\documentclass[noshowpacs, amsmath, amssymb, amsthm, aps, jmp, thmsb, twocolumn]{revtex4}
\usepackage{color,amsbsy,amssymb,latexsym,amsfonts, amsmath,cancel}
\usepackage{mathrsfs}
\usepackage{graphicx}
\usepackage{subfigure}
\newcommand{\im}{\mathrm{i}}
\begin{document}

%%%%%%%%%%%%%%%%%%%%%%%%%%%%
\title{Weak Value and the Wave-Particle Duality}

\author{Takuya Mori$^1$}
\email[Email:]{takumori@post.kek.jp}
\author{Izumi Tsutsui$^{1, 2}$}
\email[Email:]{izumi.tsutsui@kek.jp}

\affiliation{$^1$Department of Physics, University of Tokyo, 7-3-1 Hongo, Bunkyo-ku, Tokyo 113-0033, Japan\\
$^2$Theory Center, Institute of Particle and Nuclear Studies,
High Energy Accelerator Research Organization (KEK), 1-1 Oho, Tsukuba, Ibaraki 305-0801, Japan
}

\date{\today}
\begin{abstract}
The weak value, introduced by Aharonov {\it et al.}~to extend the conventional scope of physical observables in quantum mechanics,  is an intriguing concept which sheds new light on quantum foundations and is also useful for precision measurement, but it poses serious questions on its physical meaning due to the unconventional features including the complexity of its value.   In this paper we point out that the weak value has a direct connection with the wave-particle duality, in the sense that the wave nature manifests itself in the imaginary part while the particle nature in the real part.  This is illustrated by the double slit experiment, where we argue, with no conflict with complementarity, that the trajectory of the particle can be inferred based on the weak value without destroying the interference.  
\end{abstract}
%%%%%%%%%%%%%%%%%%%%%%%%%%%%

\maketitle
%\tableofcontents

%%%%%%%%%%%%%%%%%%%%%%%
%\section{introduction}
%%%%%%%%%%%%%%%%%%%%%%%

The weak value,  proposed earlier by Aharonov et al.~\cite{Aharonov} as a novel measurable quantity for an observable $A$, has been attracting much attention in recent years.  It is defined by 
\begin{eqnarray}
A_w=\frac{\langle \psi|A|\phi\rangle}{\langle \psi| \phi\rangle},
\label{eq1}
\end{eqnarray}
to a given process of transition from the initial (pre-selected) state $|\phi\rangle$ to the final (post-selected) state $|\psi\rangle$.
One of the reasons for the rise of interest is that it may provide a deeper understanding of \lq quantum paradoxes\rq\ and thereby elucidate the foundation of quantum mechanics.  The other is that the weak measurement, which is the procedure designed to obtain the weak value under negligible disturbance, can be useful for precision measurement or even for a direct measurement of quantum states (for a recent review, see \cite{Dressel-rev}).

Despite numerous studies motivated by these expectations, the physical meaning of the weak value $A_w$ remains still obscure, partly because it is complex-valued rather than real,  and also because it can become \lq anomalous\rq\ exceeding the range of the eigenvalues of $A$ \cite{Aharonov} or even \lq inexplicable\rq\ realizing the separation of physical property from its holder \cite{Cheshire, Denkmayr}.   
In this respect, it is argued that the real part of the weak value $A_w$ can be interpreted as the conditional average of $A$ pertinent to the process, while the imaginary part is related to the change of the transition probability \cite{Dressel}.   Meanwhile, we have witnessed a further puzzling phenomena involving the time-symmetric interpretation of quantum dynamics \cite{Danan}.   Quite recently, the anomalous weak value has been attributed to contextually of physical values \cite{Pusey}. 

In this paper, we point out yet another intriguing property of the weak value in connection with the wave-particle duality.  Specifically, we show that the wave nature manifests itself in the imaginary part of $A_w$ while the particle nature appears in the real part.  This is demonstrated by the double slit (gedanken) experiment, where the momentum weak value $p_w$ is directly related to the interference effect on the screen and similarly the position weak value $x_w$ to the trajectory of the particle.    The weak trajectory allows us to infer, without destroying the interference, that the particle takes either of the two classical paths from the slits when it is \lq not measured\rq\ or undisturbed, which is the situation presumed by weak measurement.  
Although this does not contradict with complementarity since the weak value is obtained for an ensemble, not for an individual particle, our result suggests the possibility of arguing both the wave and particle-nature simultaneously based on the weak value.

%%%%%%%%%%%%%%%%%%%%%%
%\section{weak value and quantum interference}
%%%%%%%%%%%%%%%%%%%%%%

We begin our discussion by showing a direct link between the imaginary part $\text{Im}\, A_w$ and interference, extending the work \cite{Dressel}.
For this, we first consider the transition amplitude
$K(\alpha) =\langle\psi|V_A(\alpha)|\phi\rangle$
between the two states $|\phi\rangle$ and $|\psi\rangle$ intervened by the unitary operator $V_A(\alpha)=\exp\left(-\im \alpha A\right)$.
The weak value is then obtained by 
\begin{eqnarray}
 A_w&=&
\im\lim_{\alpha\to0}\frac{1}{K(\alpha)}\frac{\partial K(\alpha)}{\partial \alpha}.
\label{eq3}
\end{eqnarray}
We may regard $V^\dagger_A(\alpha)|\psi\rangle$ as a family of post-selected states (ignoring the time evolution momentarily) 
and consider the variation of the transition probability.   If, during the transition, some interference effect with respect to
a basis set of intermediate states $\{|\chi_k\rangle\}$ arises, it can be argued explicitly
by inserting the completeness relation $\mathbb{I}=\sum_k |\chi_k\rangle\langle \chi_k|$ into the probability as
\begin{eqnarray}
|K(\alpha)|^2
&=&\sum_{k}\left|K_k(\alpha)\right|^2 +\sum_{j\neq k} K_k(\alpha)K_j^\ast(\alpha),
\label{eq5}
\end{eqnarray}
where
$K_k(\alpha) =\langle\psi|V_A(\alpha)|\chi_k\rangle\langle\chi_k|\phi\rangle$
is the transition amplitude via the intermediate process $k$ through the state $|\chi_k\rangle$.   It is then recognized that the first \lq diagonal part\rq\ in the r.h.s.~of (\ref{eq5}) corresponds to the classical transition, while the second \lq off-diagonal part\rq\ describes the quantum interference among the intermediate processes.

In the case of the double slit experiment, $A$ is the generator of translation on the screen, {\it i.e.}, the traverse momentum $p$ of the particle that goes through 
the slits, and $\alpha$ specifies the translation in the position of the particle along the screen.  
Once observed, the particle is ideally in the post-selected state given by the corresponding position eigenstate, and the interference reduces to the usual one, that is, the variation of the transition probability on the screen.  
The strength of interference may be evaluated by the \lq index of interference\rq\ defined by the logarithmic derivative of the off-diagonal part of (\ref{eq5}):
\begin{eqnarray}
{\cal I} :=\frac{1}{2}
\lim_{\alpha\to0}\frac{1}{|K(\alpha)|^2}\frac{\partial}{\partial \alpha} \left(|K(\alpha)|^2- \sum_{k}\left|K_k(\alpha)\right|^2\right).
\label{eq8}
\end{eqnarray}
Using the weak value $A^k_{w}= \langle \psi|A|\chi_k\rangle/\langle \psi|\chi_k\rangle$ associated with the intermediate process
$k$ which admits a formula analogous to (\ref{eq3}) with $K(\alpha)$ replaced by $K_k(\alpha)$, we obtain
{\small
\begin{eqnarray}
{\cal I} = 
\text{Im}\left(A_w -\sum_{k} \Pi_k\, A^k_{w}\right)
= \sum_{j\neq k}\text{Im}\left( A^k_{w}\frac{K_k(0)K_j^\ast(0)}{|K(0)|^2}\right),
\label{eq10}
\end{eqnarray}
}%
with $\Pi_k := |K_k(0)|^2/|K(0)|^2$ being the relative probability for the intermediate process $k$.   
This expresses the strength of interference ${\cal I}$ as the gap in the imaginary part of the weak value between the entire process and the 
average of the intermediate processes.

Note that each of the processes through $|\chi_k\rangle$ is counterfactual in the sense that it is not actually measured.  In fact, this is the crucial element of the
quantum transition and to be sharply contrasted to the classical counterpart for which all the processes are factual and $\Pi_k$ is a true probability (so that $\sum_k \Pi_k = 1$ holds, unlike the quantum case).
The last expression (\ref{eq10}) shows the rate of change in the interference in terms of quantities associated with the off-diagonal part exclusively, where one finds that  
the index ${\cal I}$ diverges for $K(0) \to 0$, that is, when the amplitudes sum up destructively.  

%%%%%%%%%%%%%%%%%%%%%%
%\section{double slit experiment}
%%%%%%%%%%%%%%%%%%%%%%

\begin{figure}[t]
\includegraphics[width=5.8cm]{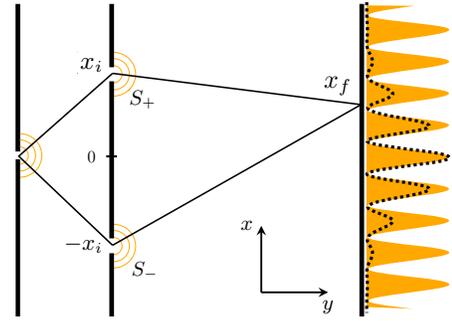}
\caption{Our simplified double slit (gedanken) experiment.  
The orange filled curve describes the transition probability when the slits are point-like, whereas the dotted curve describes the transition probability for the case where the slits are finite in size and the particle distribution is given by a Gaussian distribution around $S_\pm$. 
}
\label{QuantumInterference}
\end{figure}

We now illustrate our argument by the double slit experiment.  
Consider a particle passing through two narrow slits $S_\pm$ and later hits the screen to form an interference pattern (see FIG.\ref{QuantumInterference}).
To make our analysis simpler, we choose our state at $t=0$ by the superposition of two localized states, 
\begin{eqnarray}
|\phi\rangle=\frac{1}{\sqrt{2}} \left(|x_i\rangle+|-x_i\rangle\right).
\label{inist}
\end{eqnarray}
Our post-selection at time $t=T$ is then made by the position eigenstate at $x = x_f$,
\begin{eqnarray}
|\psi\rangle=|x_f\rangle, 
\label{finst}
\end{eqnarray}
corresponding to the point where the particle is spotted.

%%%%%%%%%
%%%%%%%%%

Ignoring the dynamics in the $y$-direction which plays no essential role in the following discussion, we just take account of the dynamics of free motion
in the $x$-direction described by the Hamiltonian,
$H={p^2}/{2 m}$,
where $m$ is the mass of the particle and $p$ the momentum along $x$. 
The time evolution is then given by the unitary transformation
$U(t)=\exp\left(-{\im H t}/{\hslash}\right)$, and for the present process the transition probability reads
\begin{eqnarray}
|\langle\psi|U(T)|\phi\rangle|^2=\frac{m}{\pi\hbar T}
\cos^2\left(\frac{m}{\hslash}\frac{x_f x_i}{T}\right).
\label{eq11}
\end{eqnarray}
This picture of interference yields undiminished fringes and is admittedly too simplistic, and perhaps one can render it more realistic by considering a Gaussian distribution for the initial state (see FIG.\ref{QuantumInterference}).  However, it will be seen that our picture is sufficient to capture the key feature of the weak value on the wave-particle duality.  

To put the present case in the general context, we introduce 
$|\phi(T)\rangle := U(T)|\phi\rangle$ and consider the family of post-selected states
$V^\dagger_p(\alpha)|x_f\rangle$ with $V_p(\alpha)=\exp\left[-\im \alpha p\right]$.  Since $p$ is the generator of 
translation, we have 
$V^\dagger_p(\alpha)|x_f\rangle=|x_f-\alpha\rangle$.  The probability amplitude is then 
$K({\alpha}) =\langle \psi| V_p(\alpha)|\phi(T)\rangle$, and the transition probability (\ref{eq11}) is just $\vert K(0)\vert^2$.
Now, the relevant completeness relation of the intermediate states is 
$\mathbb{I}=\int_0^\infty dx|x\rangle\langle x|+\int_{-\infty}^0 dx|x\rangle\langle x|$, 
according to which the transition amplitude splits as
$K({\alpha}) = {K_+}(\alpha) + {K_-}(\alpha)$,
where
${K_\pm}(\alpha)={\langle \psi|V_p(\alpha)|\phi_\pm(T)\rangle}/\sqrt{2}$
with $|\phi_\pm(T) \rangle=U(T)|\pm x_i\rangle$.

Under our choice of selections, on the screen at $t = T$ the weak value of the momentum $p_w$ and those for the partial processes $(p_\pm)_w$ are given by
\begin{eqnarray}
p_w&=&\frac{\langle \psi|\, p\, |\phi(T)\rangle}{\langle \psi|\phi(T)\rangle}
=m\frac{{x_f} +\im\, {x_i}  \tan \left(\frac{m}{\hslash}\frac{{x_f} {x_i}}{T}\right)}{T},
\label{eq18}
\\
(p_\pm)_w&=&\frac{\langle \psi|\, p\, |\phi_\pm(T)\rangle}{\langle \psi|\phi_\pm(T)\rangle}
 =m\frac{{x_f}\mp x_i}{T}.
\label{eq21}
\end{eqnarray}
Since $(p_\pm)_w$ are both real, the index (\ref{eq10}) turns out to be
\begin{eqnarray}
{\cal I} =\text{Im}\, p_w = m\frac{ {x_i}  \tan \left(\frac{m}{\hslash}\frac{{x_f} {x_i}}{T}\right)}{T}.
\label{eq22}
\end{eqnarray}
We thus see that the index ${\cal I}$ is just the imaginary part $\textrm{Im} \,p_w$, which diverges when the interference
becomes completely destructive $K(0) \to 0$ and vanishes when it is maximally constructive.

%%%%%%%%%%%%%%%%%%%%%%
%\section{weak trajectory and which path information}
%%%%%%%%%%%%%%%%%%%%%%

For a one particle system, the most tangible source of physical quantity is arguably the trajectory of the particle, so let us examine how the weak value of the position $x$
varies with time.   This is done by setting formally the pre-selected state by the retarded state $|\phi(t)\rangle = U(t)|\phi\rangle$ and the post-selected state by the advanced state 
$|\psi(t)\rangle = U(t-T)|\psi\rangle$ for $0 \le t \le T$.   The resultant weak value,
\begin{eqnarray}
x_{w}(t) := \frac{\langle \psi(t)| x |\phi(t)\rangle}{\langle \psi(t)|\phi(t)\rangle} = \frac{\langle \psi|U(T-t)xU(t)|\phi\rangle}{\langle \psi|U(T)|\phi\rangle}
\label{eq23}
\end{eqnarray}
is in general complex-valued, but it can be readily seen that, if both $|\phi\rangle$ and $|\psi\rangle$ are position eigenstates, $x_{w}(t)$ becomes real and agrees with the classical trajectory.   

\begin{figure}[t]
\includegraphics[width=7.3cm]{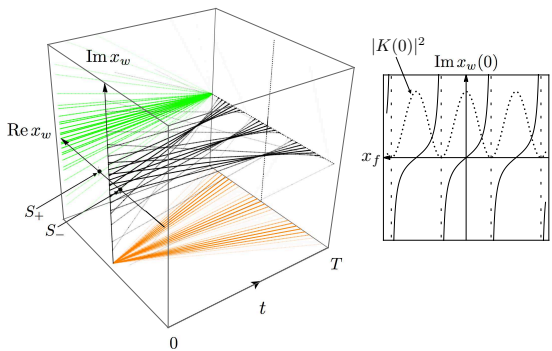}
\caption{ (Left) Weak trajectories $x_w(t)$ in the complex plane for various different post-selections with density proportional to the transition probability $\vert K(0)\vert^2 = |\langle\psi|U(T)|\phi\rangle|^2$.   The real and imaginary parts are depicted in orange and green lines and projected on the bottom and the left-back planes, respectively.
(Right) The imaginary part $\textrm{Im}\,x_w(0)$ as a function of $x_f$.   The curve diverges when the complete destructive interference occurs where $\vert K(0)\vert^2$ vanishes. 
}
\label{InterferenceDiagonal}
\end{figure}

Now, if we instead have the superposition state for $| \phi\rangle$ given in (\ref{inist}), we find
\begin{eqnarray}
x_{w}(t) = \frac{{x_f}t}{T}+\im\frac{  x_i(t-T)  \tan \left(\frac{m}{\hslash}\frac{{x_f} {x_i}}{T}\right)}{T}.
\label{eq231}
\end{eqnarray}
We then notice that 
the real part $\textrm{Re} \, x_w(t)$ corresponds to the average of the two classical trajectories coming from the two slits $S_\pm$ (see FIG. \ref{InterferenceDiagonal}).  Although this is consistent with the real part of the momentum $\textrm{Re} \,p_w$ in (\ref{eq18}), being the average ($\textrm{Re} \, x_w(0) = 0$ in particular), it cannot reasonably be regarded as a true trajectory.  In fact, this is a common feature that arises when the pre-selected state is formed by superposition, and is caused by the inability of distinction of the individual superposed states by the post-selection.  As we see shortly, this pathological behavior can be \lq cured\rq\ by rendering the distinction possible.

%%%%%%%%%
\hspace{5mm}
%%%%%%%%%

Before doing so, let us briefly discuss the imaginary part $\textrm{Im} \, x_w$, which
becomes large as the transition probability becomes small and eventually diverges when the interference
is completely destructive (see FIG. \ref{InterferenceDiagonal}).  Note that although $x$ is not
treated here as a generator for unitary transformations as $p$ is,  the connection to interference is still valid, because of the direct dynamical relation between
$\textrm{Im} \, x_w$ and $\textrm{Im} \, p_w$ obtained analogously to the Ehrenfest theorem.

The foregoing result that the real part $\textrm{Re} \, x_w(t)$ gives the average trajectory of the two classical ones
prompts us to ask what happens if we can know which of the slits the particle has gone through.  
To answer this, we bring in the spin (qubit) degrees of freedom and let the particle be in the up state $|+\rangle$ when it goes through $S_+$, and likewise in the down state $|-\rangle$ when it goes through $S_-$.  
Under this revised setup, 
our pre-selected state is given by
\begin{eqnarray}
|\phi\rangle=\frac{1}{\sqrt{2}} \left(|x_i\rangle\otimes|+\rangle+|-x_i\rangle\otimes|-\rangle\right).
\end{eqnarray}
Of course, if we perform the selection at $t = T$ by the state $|+\rangle$ or $|-\rangle$, it destroys the interference and gives nothing different from the previous setup.  However, different results arise when
we introduce, along the line of quantum eraser \cite{Scully}, an obscuring element on the \lq which path information\rq\ by adopting 
\begin{eqnarray}
|\psi\rangle=|x_f\rangle\otimes\left[\cos(\theta/2)|+\rangle+e^{\im \eta} \sin(\theta/2)|-\rangle\right]
\end{eqnarray}
for the post-selected state, which is achieved by choosing the spin up state by the measurement in the direction $(\sin\theta \cos\eta, \sin\theta \sin\eta, \cos\theta)$.
We then expect that, for $\theta=0,{\pi}$, the post-selection by $|\psi\rangle$ destroys the interference pattern as we gain the complete which path information, 
whereas for $\theta=\frac{\pi}{2}$ 
%and $\eta = 0$ 
we recover the interference fringes but lose the which path information altogether. 

To see if these expectations are realized, we introduce the spin tagged position operators, 
\begin{eqnarray}
x^\pm=x\otimes|\pm\rangle\langle \pm|,
\label{eq28}
\end{eqnarray}
which add up to 
${x^++x^-} = x\otimes\mathbb{I}$.  Obviously, these operators tell us which of the slits $S_\pm$ the particle comes from.
Then, under the free motion of the particle preserving the spin,  the tagged weak values 
$x^\pm_{w}(t)=\langle \psi(t)|x^\pm |\phi(t)\rangle/\langle \psi(t)|\phi(t)\rangle$
are found to be
\begin{eqnarray}
x^+_{w}(t)
&=&\frac{\left[x_i + (x_f-x_i)\frac{t}{T}\right]\cos(\theta/2)}{\cos(\theta/2)+e^{\im \chi}\sin(\theta/2)}
\\
x^-_{w}(t)
&=&\frac{\left[-x_i + (x_f+x_i)\frac{t}{T}\right]\sin(\theta/2)}{\sin(\theta/2)+e^{-\im \chi}\cos(\theta/2)},
\end{eqnarray}
where $\chi := \frac{2m}{\hslash}\frac{{x_f} {x_i}}{T} - \eta$.
Since $x^\pm_{w}(t)$ are both proportional to the corresponding classical trajectories,
\begin{eqnarray}
x^\pm_{cl}(t) = \pm x_i + (x_f \mp x_i)\frac{t}{T},
\end{eqnarray}
we may write 
$x^\pm_w(t) = \left[ R^\pm(\theta) +  \im I^\pm(\theta) \right]  x^\pm_{cl}(t)$
in terms of the scale factors, $R^\pm(\theta)$ and $I^\pm(\theta)$, associated with the real and imaginary parts of the weak value, respectively.   We then find, for example, the real scale factor,
\begin{eqnarray}
R^+(\theta) = \frac{1 + \cos{\theta} + \sin{\theta}\cos\chi}{2(1+\sin{\theta}\cos\chi)},
\end{eqnarray}
which has $R^+(0) = 1$, $R^+(\pi) = 0$ and $R^+(\pi/2) = 1/2$.
This shows that for $\theta=0$ the real part of the weak trajectory $\textrm{Re} \, x^+_w(t)$ coincides with the classical trajectory $x^+_{cl}(t)$ starting from the slit $S_+$ and that it vanishes for $\theta=\pi$.  This is actually expected, since $\theta=0$ ($\theta= \pi$) means that only spin up (down) particles are post-selected and hence they must come from the slit $S_+$ ($S_-$).  
Meanwhile, at the midpoint $\theta=\pi/2$ we find that $\textrm{Re} \, x^+_w(t)$ is scaled down from $x^+_{cl}(t)$ by half.  
An analogous result can be obtained for $\textrm{Re} \, x^-_w(t)$ which is proportional to the classical trajectory $x^-_{cl}(t)$ coming from the slit $S_-$ with the scaling factor $R^-(\theta)$ having $R^-(0) = 0$, $R^-(\pi) = 1$, $R^-(\pi/2) = 1/2$.  

%%%%%%%%%%%%%%%
\begin{figure}[t]
{\includegraphics[width=5.8cm]{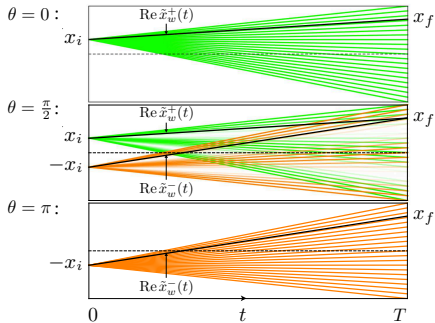}}%
\caption{
Tagged weak trajectories for the cases $\theta=0, \frac{\pi}{2}$ and ${\pi}$ with density proportional to the transition probability $|\langle\psi|U(T)|\phi\rangle|^2$.  At  $\theta=0, \pi$ where we have the perfect which path information, the interference pattern disappears and either of $\textrm{Re} \, \tilde x^\pm_w(t)$ gives one possible trajectory for the post-selection made at $x_f$.  At $\theta=\frac{\pi}{2}$ where no information on the path is gained and the interference observed, both $\textrm{Re} \, \tilde x^\pm_w(t)$ are available as trajectories from $\pm x_i$ to $x_f$.   
}
\label{RealPartOfWeakValue}
\end{figure}
%%%%%%%%%%%%%%%

The discord between the real part of the tagged weak trajectory and the classical one, appearing when the post-selection does not allow the complete which path information, is in fact an artifact arising from the behavior of the transition amplitude contained in the denominator of the weak value.  To remove this artifact, one may consider the \lq normalized\rq\ weak trajectory $\tilde x^\pm_{w}(t) := x^\pm_{w}(t)/R^\pm(\theta)$ so that $\textrm{Re} \,\tilde x^\pm_{w}(T) = x_f$ is fulfilled.  This simple adjustment at $t = T$ yields 
\begin{eqnarray}
\textrm{Re} \, \tilde x^\pm_w(t) = x^\pm_{cl}(t).
\end{eqnarray}
Thus, when we measure $\tilde x^+$ ($\tilde x^-$) weakly under the post-selection at $x_f$, the particle is surely found to have passed through $S_+$ ($S_-$), taking the classical trajectory from $x_+$ ($x_-$) to $x_f$ (see FIG.\ref{RealPartOfWeakValue}).

%%%%%%%%%%%%%%%%%%%%%%
%\section{Conclusion and Discussions}
%%%%%%%%%%%%%%%%%%%%%%

To summarize, we have shown that the imaginary part $\textrm{Im} \, A_w$ of the weak value signifies the wave nature in terms of the rate of interference.  In particular, in the context of double slit experiment the imaginary part $\textrm{Im} \, p_w$ of the momentum gives the index of interference.   In the same context, the real part $\textrm{Re} \, x_w$ of the position exhibits the particle nature through the classical trajectory, if an additional device that allows for the which-path information is equipped.  Our weak trajectory is defined purely from the position and differs from the one considered in \cite{Wiseman, Kocsis} in which a combination of velocity and position is used to obtain the dynamics of the Bohmian mechanics.

The real part of the normalized weak value of the tagged position $\tilde x_w^\pm(t)$ indicates that the particle comes either of the two slits while still yielding the interference pattern.  We emphasize that this is not in conflict with complementarity,  as the weak value is obtained for an ensemble, not for an individual particle.   Nonetheless, in view of the presumption that the weak measurement causes negligible disturbance for the particle, our result seems to suggest the na\"{\i}ve picture that, even when unseen, each of the particles in the ensemble is following the classical path before producing the interference pattern.  
It is our hope that the present study serves to gain a deeper understanding of the role played by the weak value in quantum mechanics.

\hspace{3mm}

%%%%%%%%%%%%%%%%%%%%%%
%\begin{acknowledgments}
I. T. thanks Prof. A. Wipf for useful discussions.  
This work was supported in part by the Grant-in-Aid for Scientific Research (C), No.~25400423 of MEXT, and by the Center for the Promotion of Integrated Sciences (CPIS) of Sokendai.
%\end{acknowledgments}
%%%%%%%%%%%%%%%%%%%%%%

%%%%%%%%%%%%%%%%%%%%%%%%%%%%%%%

%%%%%%%%%%%%%%%%%%%%%%%%%%%%%%%


\begin{thebibliography}{2}
\bibitem{Aharonov} 
Y. Aharonov, D. Z. Albert, and L. Vaidman, Phys. Rev. Lett. $\bold{60}$, 1351 (1988).
\bibitem{Dressel-rev} 
J. Dressel, M. Malik, F.M. Miatto N. Jordan, and R.W. Boyd, Rev. Mod. Phys. $\bold{86}$, 307 (2014).
\bibitem{Cheshire} 
Y. Aharonov, S. Popescu, D. Rohrlich, and P. Skrzypczyk, New J. Phys. $\bold{15}$, 113015 (2013).
\bibitem{Denkmayr} 
T. Denkmayr, H. Geppert, S. Sponar, H. Lemmel, A. Matzkin, J. Tollaksen, and Y. Hasegawa, Nature Commun. $\bold{5}$, 4492 (2014).
\bibitem{Dressel} 
J. Dressel  and A. N. Jordan, Phys. Rev. A $\bold{85}$, 012107 (2012).
\bibitem{Danan} 
A. Danan, D. Farfurnik, S. Bar-Ad, and L. Vaidman, Phys. Rev. Lett. $\bold{111}$, 240402 (2013).
\bibitem{Pusey} 
M.F. Pusey, \lq\lq Anomalous weak values are proofs of contextuality\rq\rq, arXiv:1409.1535.
\bibitem{Scully}
M. O. Scully, B. G. Englert, and H. Walther, Nature (London) $\bold{351}$, 111(1991).
\bibitem{Wiseman}
H.M. Wiseman, New J. Phys. $\bold{9}$, 165 (2007).
\bibitem{Kocsis}
S. Kocsis, B. Braverman, S. Ravets, M. J. Stevens, R. P. Mirin,  L. K. Shalm, and A. M. Steinberg,
Science $\bold{332}$, 1170 (2011).
\end{thebibliography}
\end{document}